\documentclass[twocolumn,showpacs,preprintnumbers,amsmath,amssymb]{revtex4}

\usepackage{graphicx}
\usepackage{dcolumn}
\usepackage{bm}

\begin{document}

\title{The Positronium state in quartz}

\author{B. Barbiellini$^1$ and P.M. Platzman$^2$}

\affiliation{$^1$ Department of Physics, 
Northeastern University, Boston, 
Massachusetts 02115, USA \\
$^2$ Bell Laboratories, Alcatel-Lucent, 
600 Mountain Avenue, Murray Hill, New Jersey 07974, USA}

\begin{abstract} 

The positronium state in quartz 
is described by a linear superposition
of two states: the first describing
the free positron in the crystal and the second
corresponding to a positronium Bloch wavefunction
in the lattice. The condition for positronium 
formation in the 
electron gas is deduced by using variational
calculations of the positron binding energy 
to the electron system.
The self annihilation parameter $\kappa$ introduced
in positron lifetime experiments can be properly justified 
by using the mixed state. 
A variational method to calculate 
$\kappa$ is proposed.

\end{abstract}


\maketitle

Positronium (Ps), 
first observed in 1951 \cite{deutsch},  
is a unique probe atom to
test bound state QED \cite{qed}. 
Ps is formed in two spin states, 75\% 
as ortho-Ps with 
parallel spins (triplet state) and 
25\% as para-Ps with antiparallel 
spins (singlet state). The energy 
difference between these spin states 
(hyperfine splitting) is about 
$8.4 \times 10^{-4}$ eV.
Para-Ps annihilates in vacuum  mainly into 
photons of 511 keV. However, 
ortho-Ps annihilates in vacuum into three photons 
in order to conserve the  spin.
The calculated lifetime in vacuum 
for para-Ps is 
$\tau_p=124$ ps while the corresponding  
value for ortho-Ps is $\tau_o=142$ ns,
which is in good agreement with recent 
experimental results \cite{rubbia}. 

Near surfaces the positron wave function overlaps 
with electrons outside the Ps. 
Therefore, the annihilation with electrons having 
an antiparallel spin 
decreases the ortho-Ps lifetime. 
This process is called pick-off annihilation 
and results in two 
photons. For instance, a quartz surface binds a ortho-Ps atom
with an attractive van der Waals interaction \cite{saniz} and
the lifetime is then reduced to about 10 ns \cite{cassidy}.

Quartz is also one of the  the rare materials, 
where Ps is delocalized in a 
Bloch state leading to peaks with negligible widths 
at reciprocal lattice vectors in the angular correlation 
of annihilation radiation of the positron annihilation (ACAR) spectrum.
Brandt {\em et al} \cite{brandt} by studying the ACAR in
quartz saw a spectacular fine structure.
Greenberger {\em et al.} \cite{greenberger}
demonstated that this structure was the manifestation
of Ps formation in the form
of a Bloch state. With the help of high resolution two dimensional 
ACAR \cite{bisson} it has been possible to clearly
extract the square modulus of the Ps Bloch
function at the $\Gamma$ state in momentum space.

Moreover, Saito and Hyodo \cite{saito} showed 
the existence of two well defined 
spin states ortho- and para-Ps by measuring 
two different lifetime components and
by introducing a self annihilation parameter
$\kappa$. Inside the quartz crystal, the ortho-Ps 
lifetime  becomes as low as $270$ ps \cite{saito} 
and the quantum mechanical interpretation of the Ps state 
inside the material becomes a controversial topic 
\cite{dupasquier}.  
In fact, in some alkali metals the positron
lifetime is above $400$ ps (i.e. much larger than the 
ortho-Ps lifetime in quartz)  and the 
positron-electron pair correlation function 
is very similar to the case of Ps in vacuum.
However, one does not observe any signatures 
of real Ps formation in alkali metals 
such as singlet-triplet lifetime splittings 
or Ps spikes in the ACAR spectra or
3-photon annihilation \cite{telegdi}.

In the quartz crystal, 
the fine structure in momentum space 
also suggests a certain decoupling between 
the Ps and the electronic gas 
surrounding it.
However, how could this condition be
met inside the quartz crystal where the Ps 
sees a considerable amount of electron density? 
Moreover, how could the self annihilation 
parameter $\kappa$ be different from zero
without violating the quantum mechanical 
principle of identical particles?

In this paper, we solve the problem of the
quantum mechanical interpretation of the
Ps state in a solid by using a mixing
of the state $\Psi_1$ representing 
the positron freely 
moving in an electron system and 
the state $\Psi_2$ of a Ps 
attached to this system.
This new state is characterized 
by a mixing
angle $\theta$, which can be 
calculated variationally.
An interesting wavefunction ansatz $\Psi_2$
for the Ps
bound state to metals has been given by Bergersen 
\cite{bergersen}. 
Unfortunately, the annihilation
rates computed from this ansatz are much lower 
than the corresponding experimental values.
However, we will see 
that $\Psi_2$ may become stable
when the electron density 
is less than the typical metallic
values.

We will start by studying the 
problem of a positron in a 
homogeneous electron gas.
This system is completely 
characterized
by the electron gas parameter
\begin{equation}
r_s = ({3\over 4\pi n})^{1/3}~,
\end{equation}
where $n$ is the electron density.
For instance, the Fermi momentum 
in atomic units is $p_F=1.92/r_s$
and the Thomas Fermi screening wave number
is  $q_{TF}=1.56/\sqrt{r_s}$ \cite{bba89}. 
Variational Quantum Monte Carlo (QMC) 
calculations can be performed by using
simulation cells containing $N=226$ electrons 
and one positron \cite{boronski,sga}.
The variational wave function in the standard
Jastrow-Slater form is given by
\begin{equation}
  \Psi_1 = D_{\uparrow} D_{\downarrow}J \varphi_+ ~,
\end{equation}
where $D_{\uparrow}$ is a Slater determinant 
formed by plane waves
for 
spin up electrons,
$D_{\downarrow}$ is a similar 
Slater determinant 
for spin down electrons,
$\varphi_+$ is the positron
wave function and $J$ is a Jastrow
factor used by Boronski~\cite{boronski}.
The wave function $\varphi_+$ is 
constant since the positron
is in the lowest energy state. 
The Jastrow factor can be factorized as
\begin{equation}
J=J_{ee}J_{ep}J_{eep},
\end{equation}
where $J_{ee}$,$J_{ep}$
and $J_{eep}$ contain the
electron-electron, 
electron-positron 
and electron-electron-positron
correlations respectively. 
The factor $J$ 
has $12$ variational parameters,
which have been optimized
using the stochastic gradient
approximation (SGA) \cite{sga,sga2}.
The positron-electron binding energy $E_B$
shown in Fig.~1 is the negative of 
the electron-positron 
correlation energy given by
\begin{equation}
 E_c= <\Psi_1|H|\Psi_1>-<\Psi_0|H|\Psi_0>,
\end{equation}
where $H$ is the Hamiltonian of the system and  
$\Psi_0$ is the state of the system non 
interacting
with the positron. 
Boronski \cite{boronski} by performing variational 
QMC calculations for $r_s$ ranging
from $2$ to $10$ has observed an anomalous behavior of 
the annihilation rates at $r_s = 6$. He suggested that
this effect is an indication of Ps formation.

To check this hypothesis, 
we have considered the trial wavefunction 
proposed by Bergersen \cite{bergersen}
\begin{equation}
\Psi_2 = {\cal A}[\Psi_{N-1}\phi],
\end{equation}
where ${\cal A}$ is the 
antisymmetrization 
operator enforcing the 
Pauli principle,
$\Psi_{N-1}$ is the wavefunction 
of the system 
without an electron 
and $\phi(r_i,r_j)$ is
the Ps wavefunction
containing variational parameters.
Similar wavefunctions have been used
to describe excitons in solids
\cite{littlewood}.
Unlike the wavefunction $\Psi_1$ 
used by Boronski, 
Bergersen's original ansatz 
neglects the role of the 
electron-electron correlation 
in the calculation
of binding energy $E_B$. 
This energy can 
parametrized by 
\begin{equation}
E_B=E_{\mbox{Ps}}\frac{r_s^{\eta}}{r_s^{\eta}+12.5},
\end{equation}
where $E_{\mbox{Ps}}$ is the Ps 
binding enegy in vacuum 
and $\eta=2.2$.
We have verified as shown in Fig.~1
that for $r_s > 6$ the positron binding enery $E_B$ given
by Bergersen is larger than $E_B$ obtained from QMC calculations
using $\Psi_1$ \cite{boronski}.
Thus, this observation indicates that Ps becomes stable 
for $r_s > 6$.
Nevertheless, a more accurate study introducing 
a Jastrow factor in Bergersen's ansatz may 
change the critical value of $r_s$. 
%
%
\begin{figure}
\begin{center}
\includegraphics[width=\hsize]{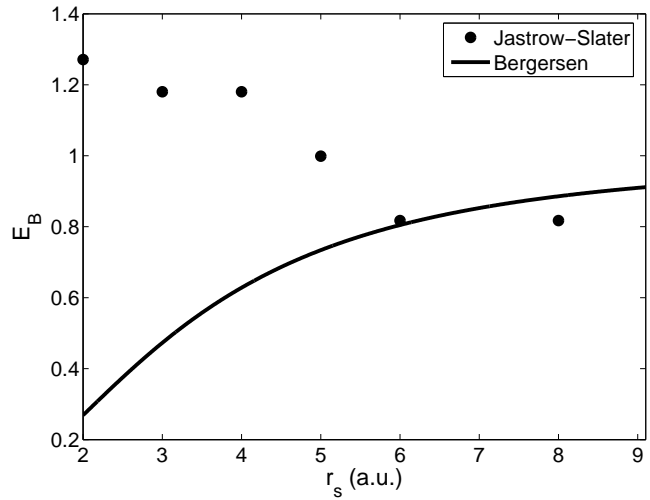}
\end{center}
\caption{Positron binding energy $E_B$ in units 
of $E_{PS}$ for the Jastrow-Slater 
and Bergersen wavefunctions 
as a function of $r_s$.
The Jastrow-Slater points are obtained from 
the QMC calculation by Boronski 
\cite{boronski}.}
\label{fig1}
\end{figure}

Next, we study the possibility in quartz 
to have a mixing
\begin{equation}
|\Psi_3>=\cos(\theta)|\tilde{\Psi_1}> + \sin(\theta) |\tilde{\Psi_2}>.
\end{equation}
The state $\tilde{\Psi_1}$ is the same 
as $\Psi_1$ except
that the Kohn-Sham orbitals for quartz 
replace the planes waves of the
homogeneous electron gas
in the Slater determinants. Moreover,
the corresponding 
positron state $\varphi_+$
is given by a periodic function 
with an energy $I$
with respect to the electron 
conduction band.
A value $I=1.7$ eV 
is obtained 
using the positron and electron chemical potentials
from density functional theory \cite{bbapos}
and the energy gap of $E_G=8.9$ eV. 
Likewise, the state $\tilde{\Psi_2}$ is the same 
as $\Psi_2$ except that the corresponding 
$\Psi_{N-1}$ is the quartz groundstate 
wavefunction
and the corresponding $\phi$ is a Ps Bloch 
wavefunction.
The variational parameters of $\Psi_3$ 
can be optimized 
by using the SGA \cite{sga,sga2}.
The mixing angle $\theta$ 
and the self-annihilation parameter $\kappa$
\cite{saito} are related via the formula
\begin{equation}
\kappa=\sin^2(\theta).
\end{equation}
The angle $\theta$ or the parameter $\kappa$ 
can also be calculated with the SGA.
This method can be used to verify
the experimental value 
$\kappa=0.34$ \cite{saito} and 
the Ps binding energy 
$4.8$ eV for quartz
\cite{nagashima}.

In conclusion, we have presented a state
representation of a Ps entangled with the electronic
structure of quartz. This model is suitable to explain
the fine structures observed in the momentum density 
distribution and the singlet-triplet positron
lifetime splitting.

We thank R. Saniz and A.P. Mills for useful
discussions. This work was supported 
by the US Department 
of Energy contract DE-FG02-07ER46352 and
benefited from the allocation of computer 
time at the NERSC and the 
Northeastern University's Advanced Scientific 
Computation Center (NU-ASCC).

\end{document}